*The following article has been submitted to the Journal of Applied Physics. After it is published, it will be found at Link.*

# Coherent Phonon Blocking in Superlattices

Anil Erol[1], Xiang Hua[1], Kirby Myers[1], Lawrence Friedman[1], Alexander Marakov[1], Melissa G. Loving[1], Joshua Shipman[1], Arsha Mamoozadeh[1], Sarah Millen, Ronald J. Warzoha[1], Jeremy Clark[1], and Robert M. Young[1]

[1]Northrop Grumman Corporation

**Abstract**

At cryogenic temperatures, phonons become one of the dominant energy carriers and thus can strongly influence the performance of electronic and sensing devices. Phonon transport departs fundamentally from the diffusive behavior typically seen at higher temperatures, as cryogenic phonons exhibit ballistic transport, propagate as waves, and possess large coherence lengths. Since phonons at millikelvin temperatures do not follow Fourier's law and are only weakly coupled to electrons, controlling and isolating phonon propagation becomes a significant challenge. Conventional approaches to thermal isolation rely on diffusion-based boundary scattering or amorphous materials, offering limited control over the spectral properties of transported phonons.

In this work, we present a wave-mechanics based framework that predicts phonon transmission and thermal resistance of arbitrarily thick superlattices while retaining all acoustic branches, mode-conversion pathways, and angles of incidence. By enforcing phase coherence, our model predicts frequency-dependent transmission through arbitrary multi-layered structures.

We use a genetic algorithm (NSGA-II) to efficiently select both materials and layer thicknesses. Our success criterion is that constituent layers satisfy the quarter-wavelength condition of the dominant phonon frequencies at a target temperature. This strategy identifies novel bilayer combinations that achieve thermal resistance of up to 3000 times greater than previously reported superlattices.

The identified superlattices are poised to advance any technology that relies on coherent acoustic scattering, from ultra-low-temperature thermal insulation in superconducting flip-chip assemblies to phonon-blocking components in micro- and nano-electromechanical systems. Our methodology thus provides both a comprehensive predictive tool and a systematic pathway for engineering next-generation phononic metamaterials.

## Introduction

Low-temperature superconductive electronics have seen a gradual rise in use as sensitive measuring devices (bolometers, NIS junction thermometers, SQUID magnetometers, transition-edge sensors, etc), low-power classical logic families (RQL, eRSFQ, AQFP Logic) and relatively easy to fabricate qubits. At sufficiently low temperatures, quantized lattice vibrations—also known as phonons—become the primary energy carrier, especially in superconductive devices and insulating materials, which creates a strong demand for controlling and isolating phonons from sensitive electronics. Device applications affected by phonon-dominated transport can be divided into two categories based on the spectroscopy of phonons needed to be blocked: (1) narrow-band and (2) broadband.

Narrow-band phonon blocking is relevant in applications where non-thermal phonons can interact with sensitive devices, either triggering false detection events or destroying the coherence of noise-sensitive states. For example, phonons at energy levels above 2Δ of a superconductor could break Cooper pairs in Kinetic Inductance Detectors (KIDs) [1], transmons, superconducting resonators and similar devices. Likewise, athermal phonons can initiate false positive events in transition-edge sensors [2] and bolometers [3]; reducing excess noise without killing thermal responsiveness would enhance detector performance. Phonon downconversion techniques and phonon traps have been designed specifically to mitigate quasiparticle loss and suppress correlated errors in superconducting qubits [4],[5].

Broadband phonon blocking remains relevant in thermally sensitive applications, requiring ultra-high thermal resistance materials. Examples of these applications include flip-chip assemblies that require thermal separation of readout chips from noise-sensitive chips (Figure 1.a); cryogenic fridge components that demand large thermal isolation to improve cooling efficiency; and thermally insulative, signal-carrying circuitry. Various devices, such as [6][7][8][9] and phononic crystals[10], have been studied to suppress the flow of phononic heat via coherent and interference-based mechanisms. These devices exploit the wave-like nature of phonons, which can exhibit strong destructive interferences when passing through quarter-wave thick layers of acoustically mismatched materials (Figure 1.b).

This study takes a fundamental departure from existing methods for designing phonon Bragg devices by explicitly optimizing for coherent phonon wave effects inside superlattice-like structures. We employ a wave-mechanics based model that accounts for all acoustic phonon modes, mode conversions, and angles of incidence with frequency-dependence. In fact, the results show that traditional means of designing laminates with single-interface models such as the acoustic mismatch model (AMM) [9][12][13] and the diffuse mismatch model (DMM) [14], or multi-layer physics models accounting for only one acoustic branch and/or normal incidence angles [15] are not sufficient for predicting phonon coherence effects in multi-layered structures.

Previous research has shown that stacked layers of differing materials can create a stop-band filter for phonons. Narayanamurti et al. [16] built a GaAs/AlGaAs superlattice and using superconductor-insulator-superconduct (SIS) junctions as phonon generators and detectors (which allow measurement of phonons with energies $h\nu \geq 2\Delta$, that is, twice the superconducting energy gap of the detector) observed attenuation at 1.05 K in the region from 0.65 to 0.8 meV in phonon energy, as expected for their layer thicknesses. Tamura et al. [15] employed an AlGa/GaAs superlattice with a pulsed laser excitation source and an SIS detector at 1.8 K, observing phonon stop-bands and imaging strong phonon focusing effects determined by favored propagation directions through the GaAs substrate. Furthermore, formation of phonon stop-bands by multi-layered structures is not limited to epitaxial crystallographic alignment of the atoms. Using SIS junctions as phonon generator and detector at 1 K, Koblinger at al. [17] built alternating amorphous Si (a-Si) and amorphous $SiO_2$ layers that exhibited strong and broad absorption bands around 210 and 420 GHz center frequencies, and a 30 period multi-layered structure incorporating hydrogen of a-Si:H alternated with amorphous silicon nitride (a-$SiN_x$:H) showing strong absorption at 570 GHz. Working at room temperature with a Si/SiGe superlattice and a femtosecond laser pulse with gated time detection, Ezzahri et al. [18] separated Brillion Zone oscillations from the thermal background and measured coherent phonon Bragg reflections centered at 283 and 527 GHz. Of particular interest to the present work, Belliard et al. [19] formed a superlattice of Mo/a-Si and measured at 15 K by optical pumping and reflection a strong and broad phonon stop-band from 0.09 to 0.18 THz.

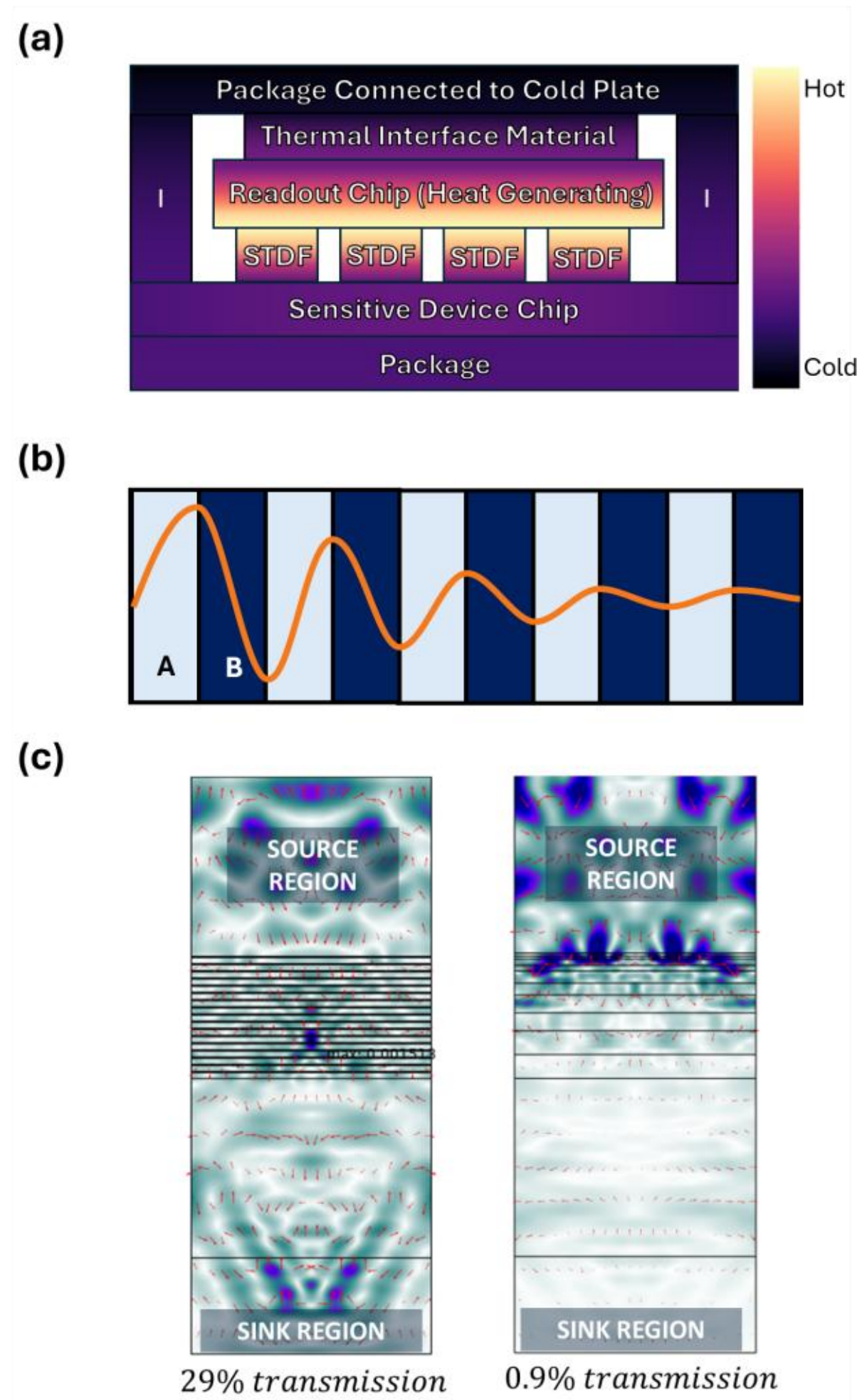

Figure 1 – *(a) Sensitive devices on flip-chip assemblies can be protected by placing standoffs with large thermal resistance, e.g., superconducting bilayers that can filter phonons. (b) Phonons can be filtered with destructive interference in quarter-wave bilayers. Materials A and B must have large acoustic mismatch, interfaces must be smooth relative to phonon wavelengths and phonon coherence must be greater than thickness of the stack. (c) Reflections of a phonon, modeled as an elastic wave in COMSOL Multiphysics®[1], are shown at time $t_1$ for two cases: (left) laminate with same bilayer thicknesses; (right) laminate with varying thicknesses. While bilayers of the same thickness can reduce transmission down to 29%, varied thicknesses can achieve 0.9% transmission by incidentally containing a combination of layers that nearly matches the quarter-wavelength of the phonon. This was an initial test case utilizing materials similar to Al and W.*

The present work offers a wave mechanics-based modeling and design framework for determining optimal superconducting laminates (i.e., superlattices) that exhibit broadband phonon filtering, yielding exceptionally large thermal resistance. The study was motivated by preliminary simulations of time-dependent elastic waves in COMSOL Multiphysics® (Figure 1.c) showing significant reduction in phonon transmission with two randomly selected, acoustically mismatched materials when altering the layer thicknesses (Figure 1.c). To achieve very large thermal resistances, our model captures these dominant wave-like phonon characteristics at ultra-low temperatures, such as large phonon coherence lengths and specular scattering mechanisms, which could result in broadband destructive interference in a laminate structure when bilayer thicknesses are varied to match a wide range of phonon frequencies. Precise layer thicknesses in this structure were determined by a custom-built coherent layered acoustic mismatch model (CLAMM), which predicts phonon transmission, wrapped inside an optimization scheme specially designed for this problem via a multi-objective genetic algorithm. The optimization's objectives were to find a set of superconducting materials and layer thicknesses that minimize phonon transmissions at a given temperature, while also minimizing total stack thickness and maximizing average $T_c$ of the stack. Only superconducting metals were considered for the optimization so that the laminate could serve both thermal isolation and electrical signal routing. Lastly, the proximity effect was modeled for one of the optimal laminate configurations to estimate the operating temperature range of the laminate.

[1] COMSOL® and COMSOL Multiphysics® are registered trademarks of COMSOL AB.

**Methods**

Phonons can propagate through a superlattice via several mechanisms depending on the size of the phonon wavelengths relative to the roughness of the interfaces and the crystal lattice. At very small wavelengths (high frequencies corresponding to high temperatures), phonons travel diffusely through the bulk material. In this work, we only consider large enough wavelengths of phonons that travel ballistically in the bulk, which dominate at low temperatures. However, even in the low-temperature regime, there are three distinct mechanisms by which ballistic phonons propagate through a multi-layered structure. The first involves diffuse scattering at the interfaces, when the interface roughness is comparable to or greater than phonon wavelengths (see Figure 2.a). In this case, phonons lose all phase information at the interface and scatter randomly in every direction. The other two cases require phonon wavelengths to be much greater than the surface roughness, which enables specular scattering that preserves phase information across the interfaces and consequently, the transmission is angle-dependent. If the phonon coherence length is smaller than the layer thicknesses, or when the layers are polycrystalline, phonons lose coherence in between interfaces (Figure 2.b); if the phonon coherence length is larger than layer thicknesses, coherence can be preserved throughout the bulk of the layers, enabling phonons to behave like waves (see Figure 2.c). This work focuses on coherent, specular scattering.

To model the phonon transmission, we utilize a coherent layered acoustic mismatch model (CLAMM), which assumes specular scattering and coherence through the layers, resulting in wave effects propagating across an entire laminate. The internal scattering of phonons in the bulk regions of the layers is neglected [20]. This model is appropriate when the interfaces are smooth compared to phonon wavelengths, and the layers' thicknesses are smaller than the phonon coherence length $\lambda_c$ . The coherence length $\lambda_c$ can be calculated by $\lambda_c = ch/k_B T$, where $c$ is the speed of sound, $h$ is Planck's constant, $k_b$ is the Boltzmann constant, and $T$ is temperature [20]. For many crystalline materials, the coherence length can be several μm at 100 milliKelvins (e.g., with a speed of sound of 5000 m/s, $\lambda_c = 2.4\ \mu m$). While this type of calculation is only applicable for bulk crystalline materials, it gives us an idea of the approximate order of magnitude of $\lambda_c$ before introducing a superlattice-like structure. Therefore, if the total superlattice is thinner than a few microns with relatively low interfacial roughness, then phonons can preserve their coherence within the entire laminate. Lastly, we only assume that heat is carried by phonons in the superconductors—thus, this model could be applicable to other materials that rely on phonons for heat transport.

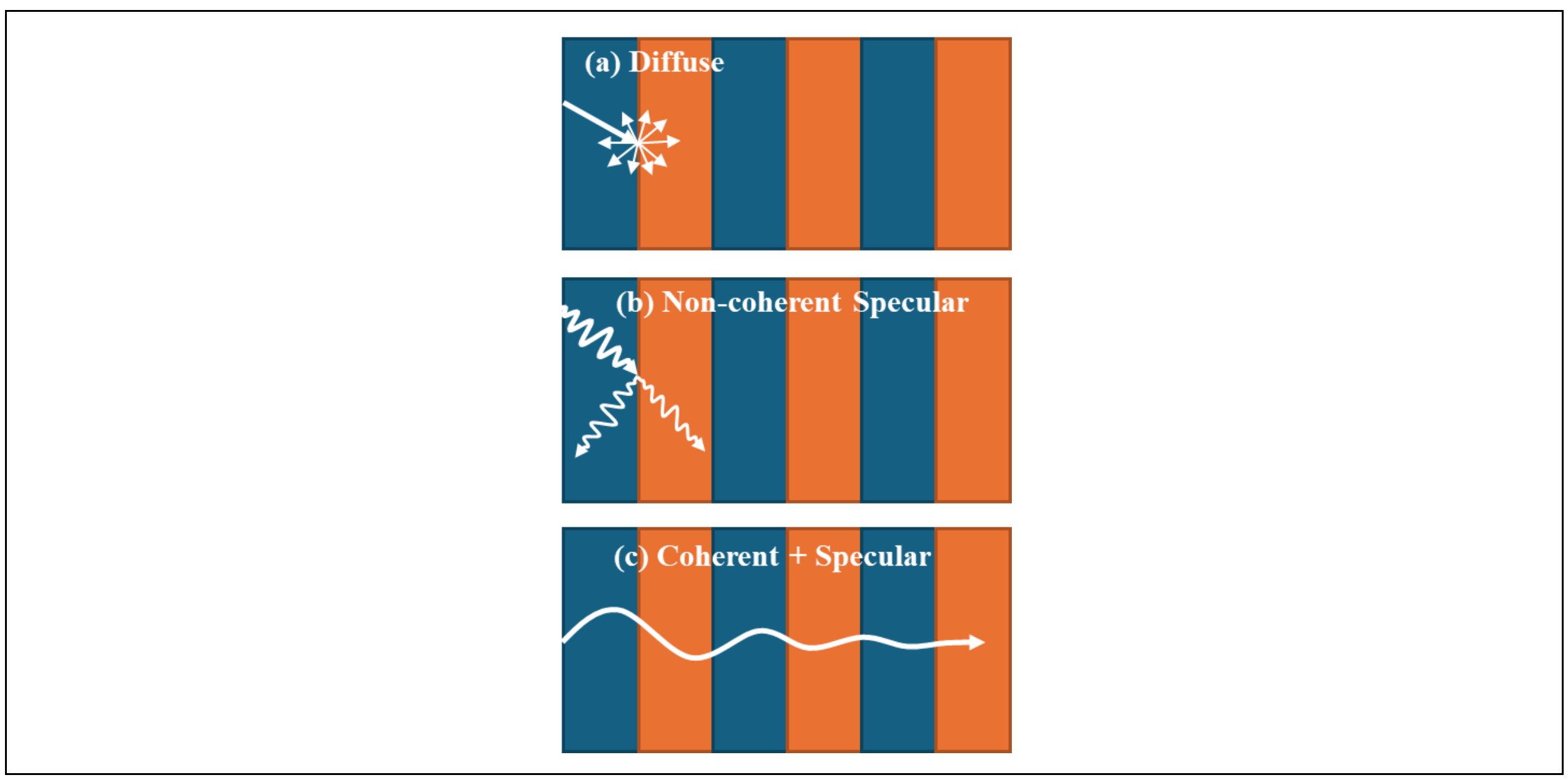

Figure 2 – *Phonon propagation through layered solid media can be described as (a) diffuse scattering at interfaces with ballistic propagation through the bulk, (b) specular scattering at interfaces with ballistic bulk propagation and (c) coherent, specular scattering where phonon characteristics are elastically coupled in both the bulk and interfaces.*

The calculation of the phonon transmission rate in the laminate system is based on the transfer matrix method for phonons [15], assuming phonons are represented by planar elastic waves. The displacement field, $\boldsymbol{u_l} = u_l\hat{\boldsymbol{\imath}} + v_l\hat{\boldsymbol{\jmath}}$, and mechanical stresses, $\sigma_{xx,l}$ (normal) and $\sigma_{xy,l}$ (shear), must be continuous at every interface. The general form of the displacements is provided by

$$\boldsymbol{u}^{(m)}(\boldsymbol{r}, t) = \hat{\boldsymbol{e}}_m\left(A_m^+ e^{i(\boldsymbol{k_m}\cdot\boldsymbol{r}-\omega t)} + A_m^- e^{-i(\boldsymbol{k_m}\cdot\boldsymbol{r}-\omega t)}\right) \tag{1}$$

where $m$ is the phonon mode (L or T), the $+$ and $-$ signs denote forward and backward propagation, while $\hat{\boldsymbol{e}}_L = (\cos\theta, \sin\theta)$ and $\hat{\boldsymbol{e}}_T = (-\sin\theta, \cos\theta)$. The unknown amplitudes $A_m^+$ and $A_m^-$ need to be solved in order to determine the total reflected and transmitted energies. With the displacements provided in (1), the strain tensor can be calculated by

$$\boldsymbol{\epsilon} = \frac{1}{2}(\nabla\mathbf{u} + (\nabla\mathbf{u})^T). \tag{2}$$

and subsequently, the stress field is given by

$$\boldsymbol{\sigma} = \lambda(\nabla\cdot\boldsymbol{u})\boldsymbol{I} + \mathbf{2}\mu\boldsymbol{\epsilon} \tag{3}$$

Note that the stress in (3) assumes isotropic elastic properties with Lame's constants $\lambda$ and $\mu$; $\boldsymbol{I}$ represents the identity tensor [21]. The displacements and stresses could thus be expressed for each layer using equations (1)-(3), such that a vector could be constructed in the form of

$$\boldsymbol{W}_l(z) = \begin{pmatrix} u_l(x) \\ v_l(x) \\ \sigma_{xx}(x) \\ \sigma_{xy}(x) \end{pmatrix} = \boldsymbol{h}_l \boldsymbol{A}_l \tag{4}$$

in which $u_l$ and $v_l$ are the x- and y- components of the displacement; $\sigma_{xx}$ and $\sigma_{xy}$ are the normal and shear stresses, respectively; $\boldsymbol{h_l}$ is a matrix with components relative to the amplitude vector $\boldsymbol{A_l}$ for layer $l$. This simple formulation allows for the writing of $\boldsymbol{W_l}(x_l) = \boldsymbol{W_{l+1}}\,(x_l)$ for every interface, constructing transfer matrices $\boldsymbol{T_i}(x_i) = \boldsymbol{h_i}(x_i)[\boldsymbol{h_i}(x_{i-1})]^{-1}$, and solving backwards for the waves in the source domain,

$$\boldsymbol{W}_N(x_N) = (\boldsymbol{T}_N \boldsymbol{T}_{\boldsymbol{N-1}} \dots \mathbf{T_2 T_1})^N\, \boldsymbol{W_1}(0). \tag{5}$$

Given (5), the amplitudes of the reflected waves in the source material could be calculated, which can be used to calculate the energy fluxes and the transmission probability $\alpha(\theta,\omega)$,

$$\alpha(\theta,\omega) = 1 - \frac{|S_{1L}|}{|S_0|} - \frac{|S_{1T}|}{|S_0|}. \tag{6}$$

The fluxes are defined by $S_m = \frac{\omega^2 \rho c_m}{2} |A_m^2| Re(\cos\theta_m)$ for mode $m$. The incident wave's flux is given by $S_m = \frac{\omega^2 \rho_0 c_m}{2} \left|A_0^2\right| Re(\cos\theta_0)$ [12].

The phonon transmission coefficient per mode m, $\Gamma_\mathrm{m}(\omega)$, can be calculated by averaging over the angles of incidence $\theta$ from 0 to $\frac{\pi}{2}$,

$$\Gamma_m(\omega) = \int_0^{\frac{\pi}{2}} \alpha(\theta,\omega) \sin\theta_m \cos\theta_m \, d\theta_m. \tag{7}$$

The total transmission coefficient $\Gamma_{tot}^{1\to2}$ can be obtained from the weighted average of all three modes $\Gamma_j$ ($\Gamma_{tot}^{1\to N} = \sum_j c_{1,j}^{-2} \Gamma_{1,j}^{1\to N}$) and the expectation value of the transmission coefficient at a temperature $T$ is given by

$$\Gamma_{tot}^{1\to N}(\omega,T) = \int_0^{\omega_D} \int_0^{\frac{\pi}{2}} u_E(\omega,T) \alpha^{1\to N}(\theta,\omega) \sin\theta_m \cos\theta_m \, d\theta_m d\omega. \tag{8}$$

The function $u_E(\omega,T)$ represents the spectral strain energy density function of phonons [22]. After the transmission coefficient is calculated, the thermal resistance from CLAMM can be determined by

$$R_{\boldsymbol{CLAMM}} = \frac{2\left(1 - \Gamma_{tot}^{1\to \boldsymbol{N}} - \Gamma_{tot}^{\boldsymbol{N}\to 1}\right)}{2.04 \times 10^{10} \Gamma_{tot}^{1\to N}} T^{-3} \tag{9}$$

where the speeds of sound, $c_{i,j}$, for phonon modes $j = L, T_1, T_2$. Note that there is an addition to the original AMM, $1 - \Gamma_{tot}^{1\to \boldsymbol{N}} - \Gamma_{tot}^{\boldsymbol{N}\to 1}$, which corrects for the imaginary boundary paradox [6].

The primary opportunity to improve the laminate thermal resistances lies in the coherence of phonons, which gives rise to wave-like phenomena. To achieve higher thermal resistance, it is essential to optimize the layers such that phonons destructively interfere as they travel through the laminate. These Bragg-like effects depend on (1) material properties of all constituent layers, which determine phonon wavelength, and (2) the thickness of each layer, which determines the associated quarter-wavelength needed for destructive interference.

As a test case, the present study applied the theoretical framework to the design optimization of a superconducting laminate that serves as a thermally resistive standoff between two chips in a flip-chip bonded assembly. We chose a laminate configuration with twenty layers of arbitrary materials and layer thicknesses—thus, the optimizer would need to select both the material and the thickness of all twenty layers. Three design considerations were chosen to serve as independent objective functions: (1) the thermal phonon transmission of the laminate $\Gamma_{T^*}$, which is the total transmission coefficient $\Gamma_{tot}$ weighted by the spectral energy density of phonons at a target temperature $T^*$, (2) the averaged superconducting transition temperature $T_c^*$, and (3) the total laminate thickness $t_{tot}$.

The aim of the optimization routine was to either minimize or maximize the objective functions. The thermal phonon transmission $\Gamma_{T^*}$ was selected as a surrogate for the thermal resistance; as $\Gamma_{T^*}$ decreases, the thermal resistance increases. Therefore, the optimization aimed to minimize $\Gamma_{T^*}$. The superconducting transition temperature $T_c^*$ was chosen since electronic thermal conduction can dominate in superconductors well below $T_c$ due to the small but non-negligible population of quasiparticles. To completely eliminate the thermal dominance of the quasiparticles, the superconductor must be close to $T = T_c/9$ (see Appendix D for more information). Therefore, $T_c^*$ was maximized by the optimizer. Lastly, the total thickness $t_{tot}$ was selected to be minimized by the optimizer since there are often strict thickness/gap constraints in a flip-chip assembly—ideally, the laminate should be as thin as possible. Due to the large design space and the presence of three objective functions, an evolutionary algorithm (NSGA-II [23]) was selected as a multi-objective global optimization method. Details of the optimization can be found in Appendix B. Based on fabrication considerations (e.g., preserving interfacial quality), layer thicknesses were constrained to 50 nm or less per layer.

Table 1. The material properties used in this study. Note that these material properties were approximated from known elastic properties, or estimated based on existing literature.

| Material Name | $T_c\ (K)$ | $c_L\ \left(10^3\,\frac{m}{s}\right)$ | $c_T\ \left(10^3\,\frac{m}{s}\right)$ | $\rho\ \left(\frac{g}{cm^3}\right)$ |
|---|---|---|---|---|
| Al | 1.2 | 6.24 | 3.04 | 2.73 |
| Nb | 9.7 | 4.74 | 2.08 | 8.6 |
| NbN | 14 | 6.92 | 3.7657 | 8.2 |
| Re | 1.7 – 2.4 | 5.45 | 2.91 | 21 |
| Ta | 4.5 | 4.17 | 2.08 | 16.69 |
| TiN | 5.6 | 10.6 | 6.2 | 5.22 |
| V | 5.38 | 6.02 | 2.77 | 6.11 |
| Mo | 0.92 | 6.25 | 3.35 | 10.2 |
| Zr | 0.55 | 4.65 | 2.25 | 6.52 |
| Ti | 0.3 – 0.4 | 6.07 | 3.05 | 4.5 |

The complete list of superconductors allowed in the optimization are listed in Table 6. Having a combination of high and low $T_c$ superconductors allowed the search to find a variety of laminate designs that could be used in different temperature ranges.

The design optimization was conducted in two phases: (1) a broad multi-objective global search that allowed the material of any layer to change, and (2) a narrower single-objective optimization that only used a single type of bilayer. The first stage allowed us to select the best options of superconducting materials, while the second stage removed some constraints on total thickness and $T_c$ to allow the optimizer to pick the best thicknesses for a laminate built from two specific superconductors. Furthermore, these results were compared to the commonly studied superconducting bilayers made out of Nb/TiN.

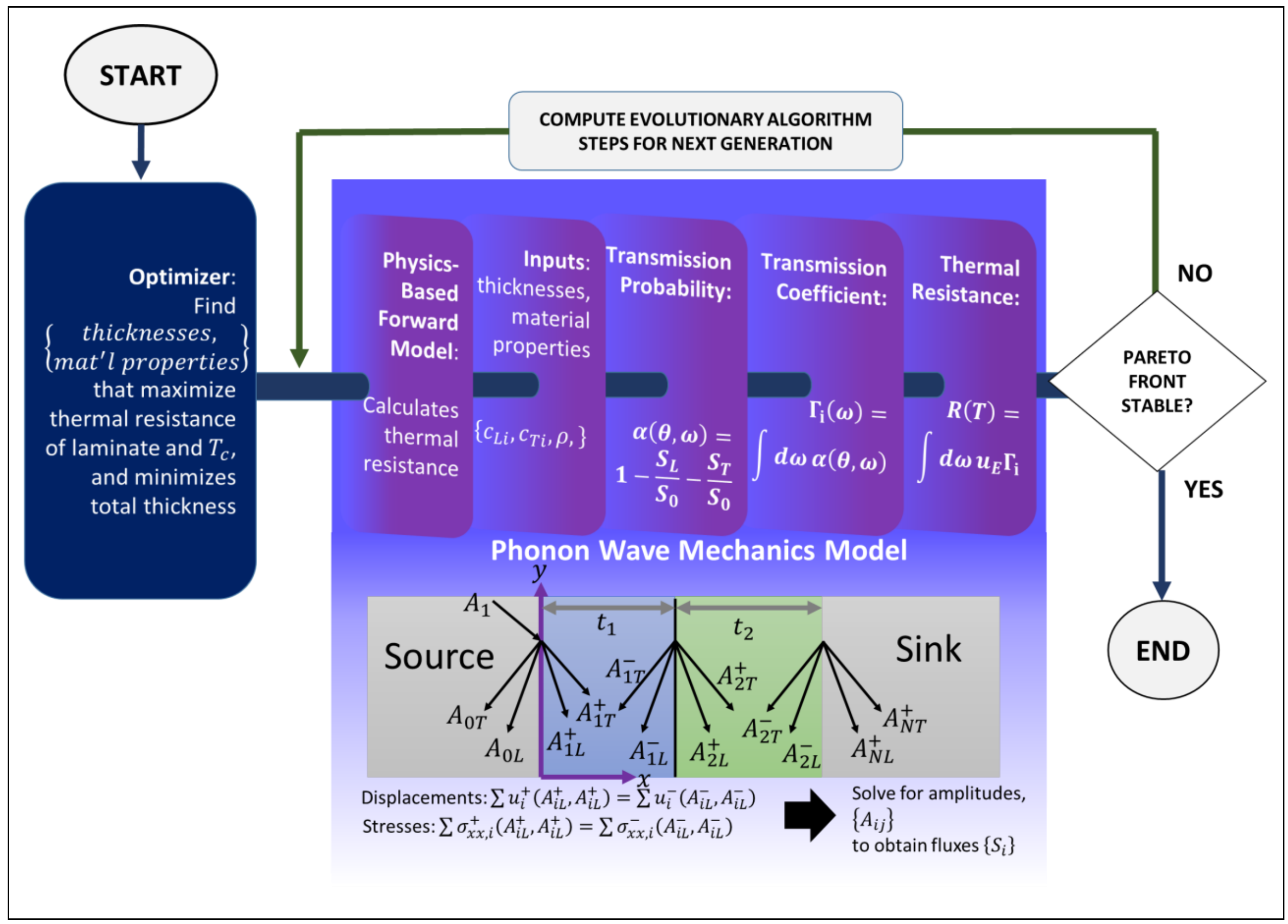

Figure 3 – *A genetic algorithm was deployed for a multi-objective optimization that sought a maximum thermal resistance of the laminate R, maximum superconducting critical temperature $T_C$, and minimum total laminate thickness.*

**Results**

The multi-layer phonon transmission model was compared to experimental results found in the literature [19] to validate the predictive capability of the transmission probability calculations. The dataset is for a superlattice consisting of amorphous silicon and molybdenum bilayers. A picosecond ultrasonics technique was used to generate acoustic phonons and the transmission was measured. Figure 4 plots the calculated transmission contours versus angle of incidence and frequency of a phonon hitting the superlattice, along with the measured transmission spectroscopy from the experimental study. Note that dark regions in the contour between 0.075 THz and 0.2 THz indicate very low phonon transmissions, which matches the trends seen in the measurements. The prior study expected more bandgaps, such as in the 0.25

THz to 0.3 THz band, by modeling phonon propagations with only normal incidence (zero degree angle of incidence). Our model shows that although there is a bandgap for normal incidence phonons between 0.25 THz and 0.3 THz, there are still channels of transmission in this frequency band when phonons enter the superlattice at around 30 to 40 degree angles of incidence. This result highlights the importance of solving for all angles of incidence when designing acoustic Bragg devices.

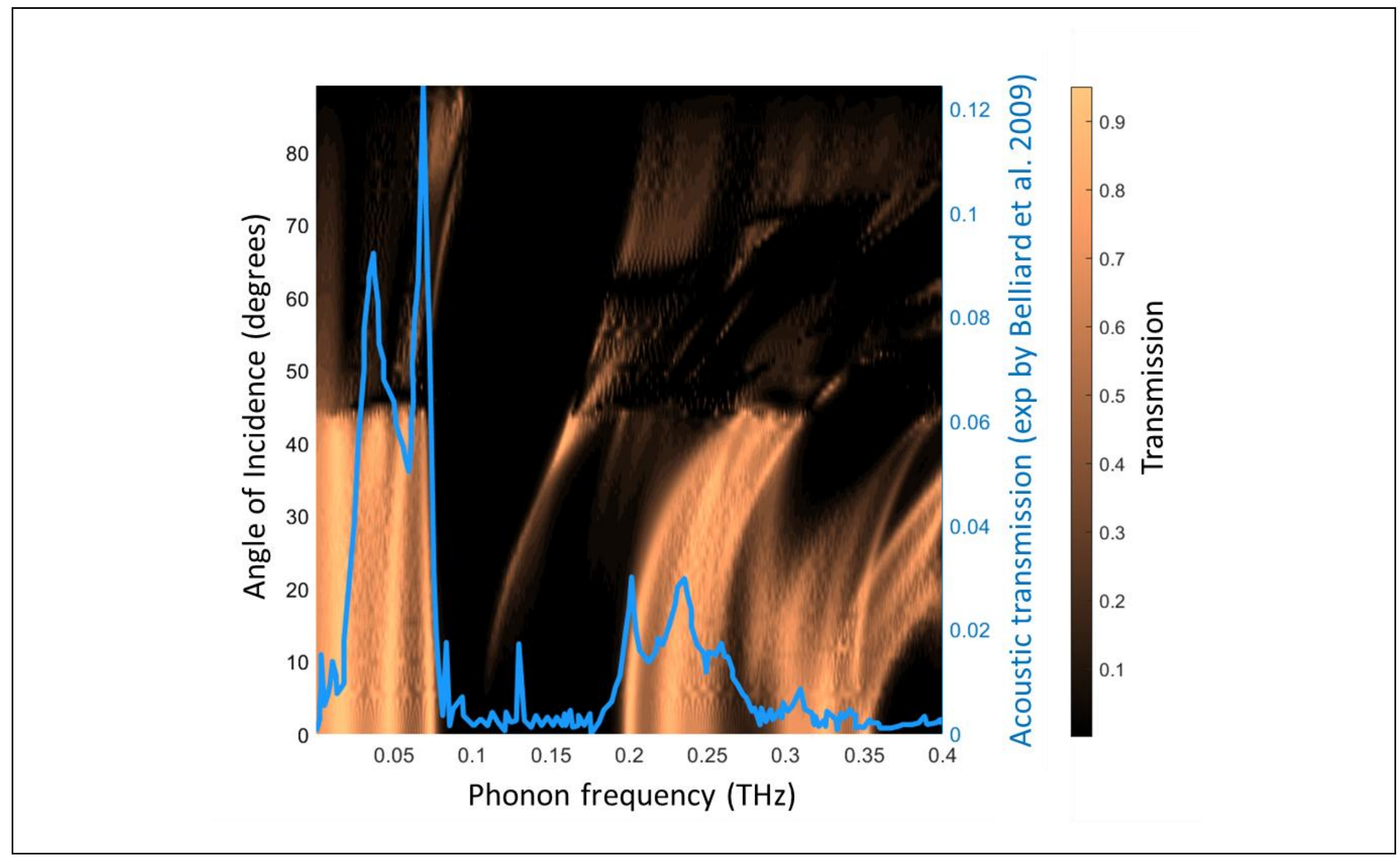

Figure 4 – *The contours plot transmission probability (from 0 to 1) calculated by the model across all angles of incidence and up to 0.4 THz in phonon frequency for a 3 bilayer superlattice consisting of amorphous silicon and molybdenum. This superlattice was tested in the literature—the measured transmission spectroscopy is plotted versus frequency by the blue curve [19].*

To build intuition for the model behavior, we predicted the phonon transmission characteristics of two superlattices made of different bilayer materials: (1) Nb and TiN, and (2) Re and Al. The Nb-TiN superlattice has been studied in recent years as a potential candidate for efficiently blocking phonons due to the large acoustic mismatch between Nb and TiN [8]. Both the original acoustic mismatch model (AMM) and the diffuse mismatch model (DMM) predict large thermal resistances for the interface between Nb and TiN: 7.6 $cm^2K^4/W$ and 18 $cm^2K^4/W$, respectively [14]. Meanwhile, our optimization consistently identified Re-Al bilayers as exhibiting exceptionally large phonon attenuation compared to other bilayer combinations. Surprisingly, the predicted thermal resistances for the Re-Al interface, 6.5 $cm^2K^4/W$ for AMM and 7.7 $cm^2K^4/W$ for DMM, are lower than the Nb-TiN interface. According to these single-interface models, Nb-TiN should perform at least as well as, if not better than Re-Al as a phonon blocker. Based on this comparison, we could develop intuition on how well single-interface models such as AMM and DMM, which engineers can readily compute, can predict the ability of a bilayer to filter out phonons as an acoustic Bragg device.

Figure 5 compares the phonon transmission coefficient for a one-bilayer and five-bilayer laminate made of either Nb-TiN or Re-Al. The layer thicknesses were chosen so that they matched the quarter-wavelength of 10 GHz transverse phonons (note: the majority of thermal phonons exist in transverse modes due to the density of states being proportional to $c_m^{-2}$ for the speed of sound in mode $m$)—this corresponds to 52 nm for Nb, 155 nm for TiN, 73 nm for Re and 73 nm for Al. By using quarter-wavelength thicknesses, these bilayers would be ideal for blocking one frequency, enabling a fair comparison between the two types of bilayers. The transmission coefficients for both one-bilayer laminates were relatively large, with the Re/Al bilayer transmission about three times lower near 10 GHz compared to Nb/TiN. However, the differences were quite drastic when comparing the five-bilayer configurations. While the five-bilayer Nb/TiN laminate showed only a marginal reduction in transmission near 10 GHz, the five-bilayer Re/Al laminate exhibited a transmission that was several orders of magnitude reduced compared to its one-bilayer configuration.

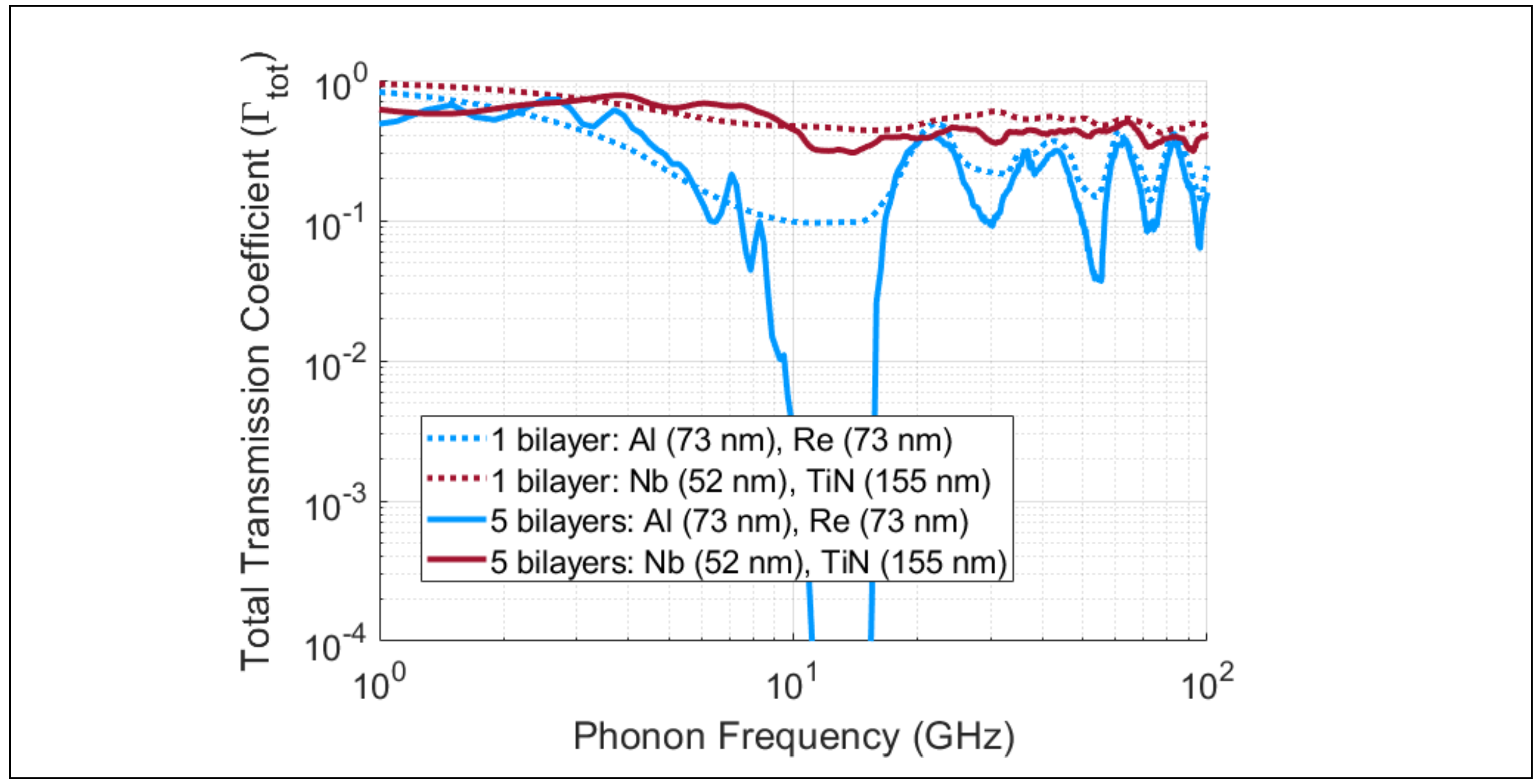

Figure 5 – *Phonon transmission coefficients are plotted for two laminates made of either (1) Re and Al or (2) Nb and TiN. Each type of laminate's transmission coefficient is plotted with 1 bilayer and 5 bilayers to show the effects of adding more bilayers.*

Next, we investigated the underlying mechanisms for why a material combination such as Nb and TiN did not benefit from adding bilayers, while materials such as Re and Al drastically did. Namely, we compared two types of ten-bilayer laminates: (1) a laminate consisting of Nb (50 nm) and TiN (10 nm) based on a recent study[8] and (2) Re and Al bilayers with varying thicknesses determined by a single-objective optimization that only varied the thicknesses to minimize the phonon transmission (see Table 2for layer thicknesses). The source and sink materials utilized the same materials as the laminate—thus, phonons entered the Nb/TiN laminate from Nb and exited into TiN; likewise, phonons entered the Re/Al laminate from Re and into Al.

Table 2. The layer thicknesses of a laminate that was generated from a single-objective optimization only considering Al and Re layers. The source material was Re and sink material was Al. The optimization was performed for blocking phonons between 10 GHz and 100 GHz. All layer thicknesses are in nanometers (nm).

| Al | Re | Al | Re | Al | Re | Al | Re | Al | Re | Al | Re | Al | Re | Al | Re | Al | Re | Al | Re |
|---|---|---|---|---|---|---|---|---|---|---|---|---|---|---|---|---|---|---|---|
| 42 | 18 | 37 | 49 | 6 | 40 | 44 | 6 | 49 | 50 | 17 | 25 | 47 | 45 | 47 | 44 | 11 | 30 | 19 | 50 |

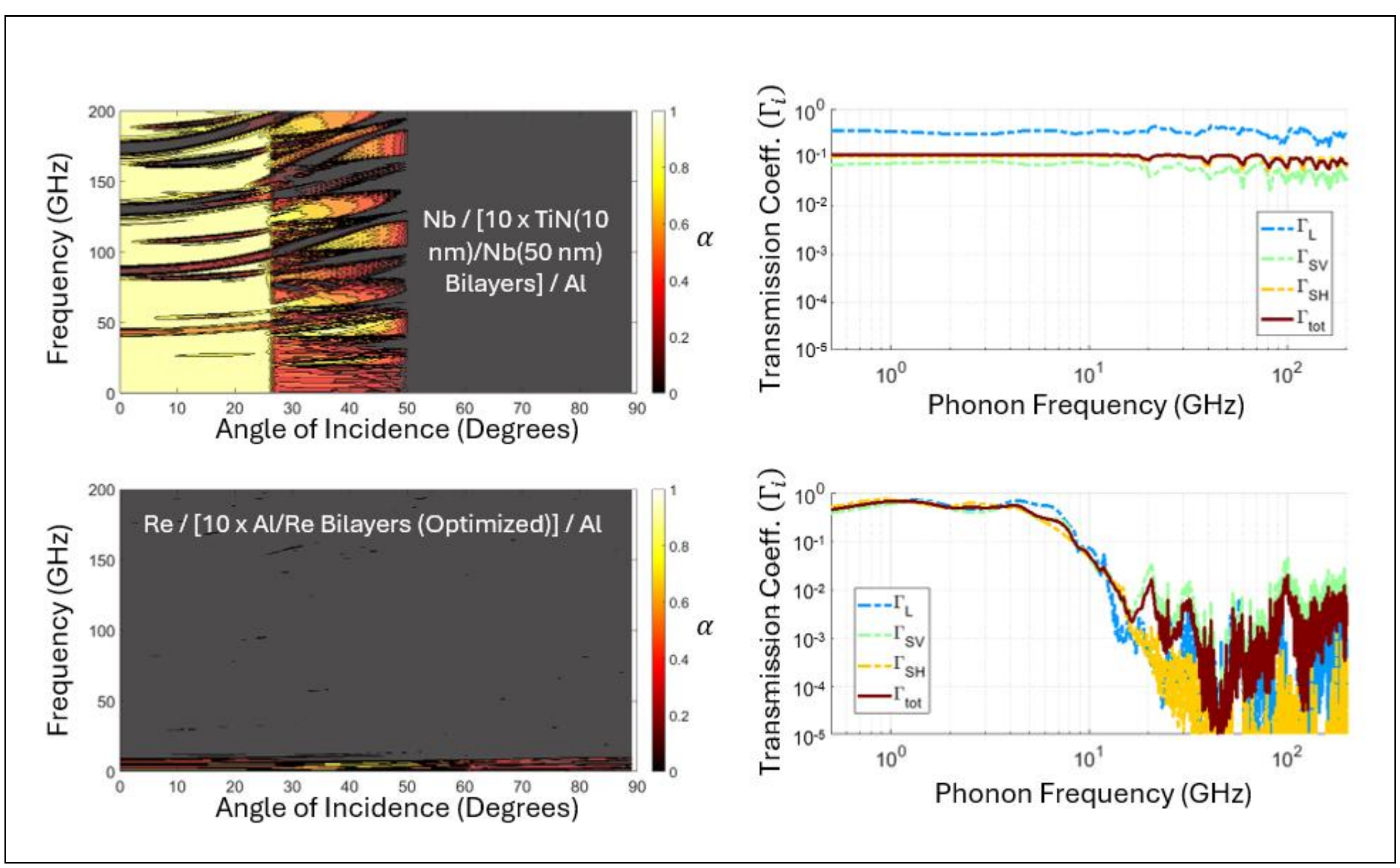

Figure 6 – *Phonon transmission probabilities and coefficients are plotted for two laminates: (a,b) 10 bilayers of Nb/TiN with varying thicknesses, (c,d) 10 bilayers of Al/Re with varying thicknesses. The transmission coefficients of individual phonon modes (L: longitudinal, SV: shear vertical, SH: shear horizontal—SV and SH correspond to transverse modes) and a total quantity are plotted. The total transmission is averaged by weighting with the density of states of each mode and normalizing by the total density of states.*

The modeling results for these two laminates are shown in Figure 6. Plotting the transmission probability of the two laminates as contour plots against both frequency and angle of incidence, as shown in Figure 6.a and Figure 6.c, helps explain how and why some material combinations can perform better than others when designing laminates as phonon Bragg devices. At the lowest frequencies, the behaviors of the laminates converge toward a single interface of either Nb and TiN or Re and Al due to the boundary conditions chosen for this comparison. For instance, the transmission rate of Nb/TiN at low frequencies matches a single interface of Nb and TiN, which has a critical angle of incidence (angle at which total internal reflection occurs) of about 50 degrees. By contrast, the transmission rate through Re/Al at low frequencies has a very high critical angle near 90 degrees (due to the similar speeds of sounds in Re and Al). The total transmission coefficients of each laminate at low frequencies tell a similar story: transmission through Nb/TiN of phonons below 1 GHz is about 0.1 while it varies between 0.4 to 0.7 for Re/Al. Thus, after accounting for the speeds of sound of the materials, we obtain similar AMM values for both single interface thermal resistances. However, when looking at the rest of the spectrum, it is clear that Re/Al bilayers significantly filter out more phonons at higher frequencies when the wavelengths are comparable to the layer thicknesses, while Nb/TiN bilayers perform similarly as a single interface. This can best be explained by the single interface behavior: while the critical angle is low, the open channels of transmission have very high rates near 1. Therefore, adding more layers does not improve phonon-blocking for Nb and TiN bilayers. By contrast, even though a single interface of Al and Re allows larger angles of incidence for phonons to pass through, the transmission rates are relatively low, ranging between 0.1 and 0.8—i.e., more reflections occur. As a result, destructive interference is better enabled in the presence of larger reflections in Re/Al bilayers at those angles. Therefore, when selecting materials for bilayers, one method could be to investigate how the single-interface behaves and whether it promotes non-negligible reflections at all angles of incidence.

Table 3. The layer thicknesses of a laminate generated by a single-objective optimization only considering Al and Re layers. Both source and sink materials were silicon. The optimization was performed for minimizing phonon propagation between 10 GHz and 100 GHz. All layer thicknesses are in nm.

| Al | Re | Al | Re | Al | Re | Al | Re | Al | Re | Al | Re | Al | Re | Al | Re | Al | Re | Al | Re |
|---|---|---|---|---|---|---|---|---|---|---|---|---|---|---|---|---|---|---|---|
| 28 | 50 | 3 | 50 | 50 | 5 | 45 | 46 | 14 | 32 | 18 | 9 | 41 | 48 | 48 | 50 | 30 | 11 | 25 | 42 |

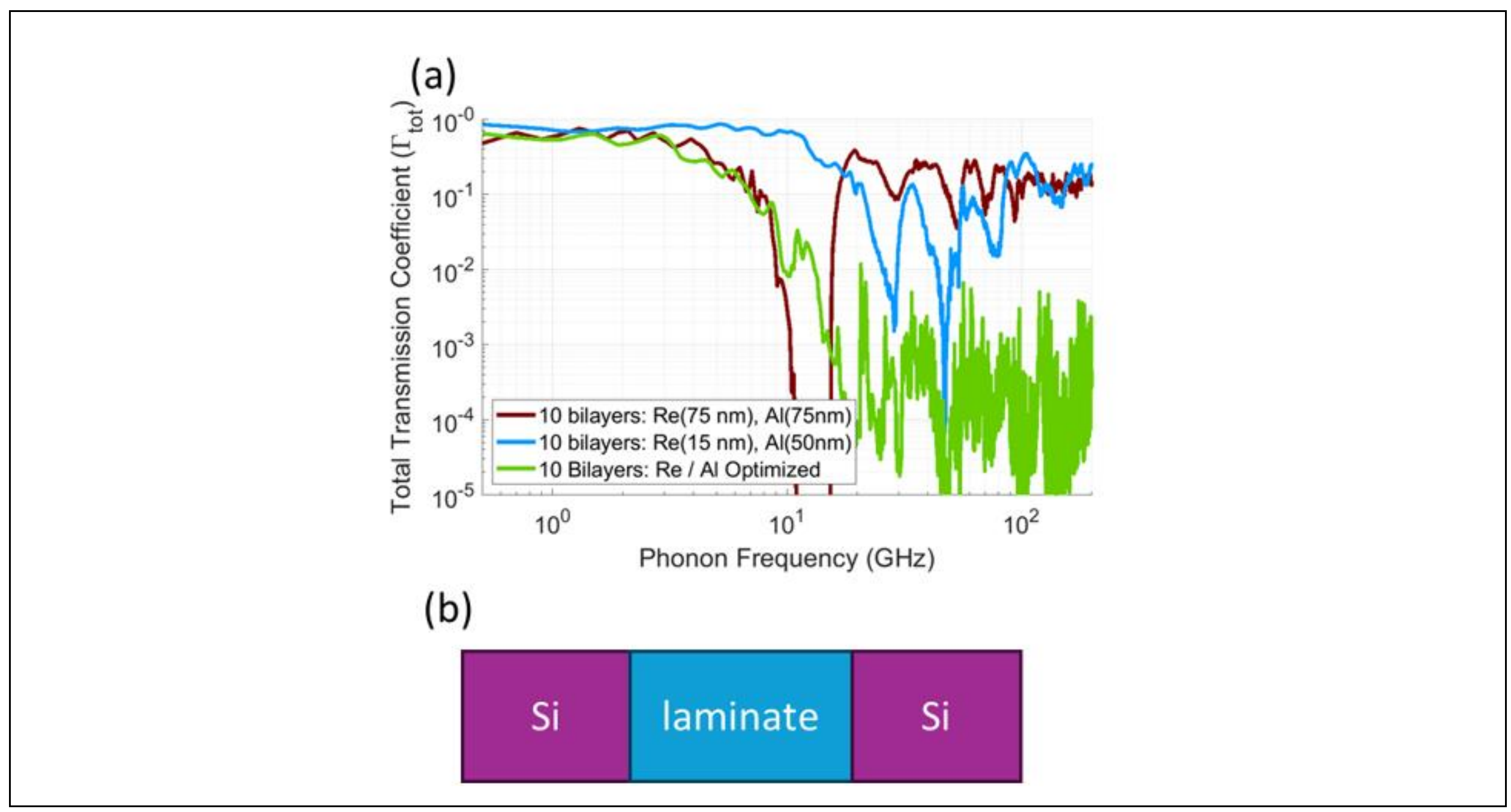

Figure 7 – *(a) Re/Al-based laminate total transmission coefficients versus phonon frequency. All three configurations contained ten bilayers (totaling twenty layers of repeating Re and Al), with the source and sink materials kept as silicon, as shown in (b).*

Next, we studied a comparison specifically for Re/Al-based laminates, as shown in Figure 7, which plots the spectral phonon transmission coefficients for three 10-bilayer laminate configurations. All three samples were simulated with phonons injecting from silicon on the source side and deposited into silicon on the sink side. Furthermore, the only difference between the configurations was the layer thicknesses. The first configuration was chosen to have layer thicknesses of 75 nm for both Al and Re to target a narrow frequency band. The second configuration consists of 15 nm of Re and 50 nm of Al, as an example of randomly selected thicknesses. The third configuration consists of varying layer thicknesses obtained from the single-objective optimization on a laminate only allowed to contain Al and Re bilayers (same as Table 2). Results demonstrated that a consistent layer thickness—75 nm in the case of Al and Re—was extremely efficient in reducing phonon transmission in a narrow band (10-15 GHz). Picking layer thicknesses arbitrarily, as in the case of the Re(15 nm)/Al(50 nm) case, was much less effective in blocking any single band as well. The optimized result, which consisted of various thicknesses optimized to block a range of phonons between 10 and 100 GHz, even with constraint on the maximum layer thickness set to 50 nm was able to achieve low transmission rates hovering around $10^{-4}$ over a band covering an entire order of magnitude (10-100 GHz).

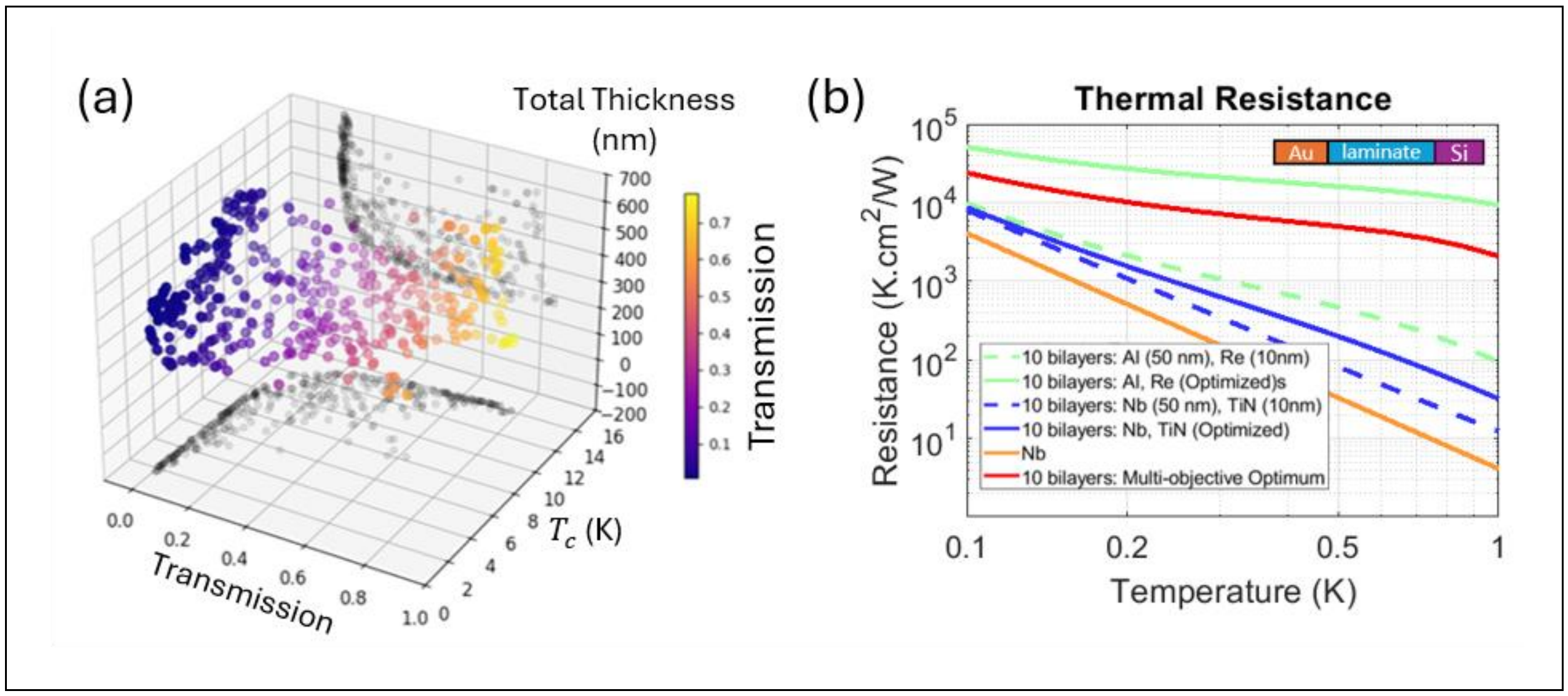

Figure 8 – *(a) Pareto front resulting from the multi-objective optimization, allowing for variations in materials and thicknesses in each layer. (b) Modeled thermal resistances of the following laminates: 10 bilayers of Al/Re, optimized vs non-optimized (constant thicknesses: 50 nm of Al / 10 nm of Re); 10 bilayers of Nb/TiN, optimized vs unoptimized (constant thicknesses: 50 nm of Nb / 10 nm of TiN); a single layer of Nb; a 10-bilayer design from the multi-objective optimization (D1 in Table 5). All optimized samples were targeting a temperature of about 800 mK. All laminates were modeled with source and sink materials as Au and Si, respectively, to match the experimental conditions of a prior study[8].*

The multi-objective optimization produced the Pareto front shown in Figure 8. The designs were allowed to have any of the superconductors listed in Table 1, and layer thicknesses allowed from 1 nm to 50 nm. Based on the application requirements, one may select across a range of options: for example, if thermal resistance is the highest priority, then the best designs are dominated by Al and Re, such as in D1 (Table 5). If the highest priority is to achieve high average $T_c$, then the best designs could be D2 or D3 (Table 5), which have high contents of NbN ($T_c = 14\ K$). Furthermore, D2 had a total thickness of only 87 nm, significantly less than the 429 nm for D1 and 236 nm of D3.

The corresponding thermal resistance of these selected designs were plotted in Figure 8.b, to compare against some of the other, previously discussed laminates in this study. These results reflect the large impact that having Re/Al bilayers have on the thermal resistance. The largest thermal resistance was once again the design that was chosen from the single-objective optimization only considering 10 bilayers of Re/Al. However, it is worth noting how important the layer thicknesses were, as the Re/Al laminate (10 bilayers) consisting of uniform thicknesses of 50 nm of Re and 10 nm of Al performed significantly worse than the optimized Re/Al laminate (by about two orders of magnitude).

Table 4. A few laminate designs selected from a single-objective optimization only considering Nb and TiN layers. The source and sink materials were gold and silicon, respectively. The optimization was performed for minimizing phonon propagation 800 mK phonons. All layer thicknesses are in nm.

| Nb | TiN | Nb | TiN | Nb | TiN | Nb | TiN | Nb | TiN | Nb | TiN | Nb | TiN | Nb | TiN | Nb | TiN | Nb | TiN |
|---|---|---|---|---|---|---|---|---|---|---|---|---|---|---|---|---|---|---|---|
| 2 | 30 | 33 | 49 | 9 | 47 | 38 | 50 | 15 | 50 | 44 | 43 | 22 | 50 | 48 | 33 | 6 | 27 | 22 | 50 |

Table 5. A few laminate designs selected from the multi-objective optimization. Source and sink materials were gold and silicon, respectively. The optimization was performed for minimizing phonon propagation between 10 GHz and 100 GHz. All layer thicknesses are in nm.

| | **D1** | | **D2** | | **D3** | |
|---|---|---|---|---|---|---|
| $\boldsymbol{f_1}$ | 0.00594 | | 0.765 | | 0.114 | |
| $\boldsymbol{f_2}$ **(K)** | 2.01 | | 13.7 | | 10.5 | |
| $\boldsymbol{f_3}$ **(nm)** | 429 | | 87 | | 236 | |
| **Layer** | **Material** | **Thickness (nm)** | **Material** | **Thickness (nm)** | **Material** | **Thickness (nm)** |
| 1 | Re | 45 | Nb | 2 | Re | 16 |
| 2 | Ta | 7 | NbN | 3 | Al | 5 |
| 3 | Al | 28 | NbN | 2 | Al | 8 |
| 4 | TiN | 15 | NbN | 6 | NbN | 8 |
| 5 | Al | 14 | NbN | 4 | V | 9 |
| 6 | V | 6 | NbN | 10 | NbN | 15 |
| 7 | Al | 17 | NbN | 7 | NbN | 9 |
| 8 | Re | 19 | NbN | 7 | Nb | 7 |
| 9 | Ta | 3 | NbN | 3 | Al | 2 |
| 10 | Al | 32 | NbN | 3 | NbN | 33 |
| 11 | Re | 36 | NbN | 2 | NbN | 32 |
| 12 | Re | 15 | NbN | 5 | Zr | 2 |
| 13 | Al | 3 | NbN | 5 | NbN | 15 |
| 14 | Re | 9 | NbN | 2 | NbN | 7 |
| 15 | Re | 41 | V | 2 | NbN | 10 |
| 16 | Al | 26 | NbN | 8 | NbN | 2 |
| 17 | Ta | 28 | NbN | 8 | Re | 6 |
| 18 | Re | 21 | NbN | 3 | Al | 3 |
| 19 | Al | 20 | NbN | 2 | NbN | 32 |
| 20 | Re | 44 | NbN | 3 | Al | 14 |

Figure 8.b shows the temperature-dependent thermal resistance of the optimized Al/Re laminate alongside four reference curves: (1) the thermal resistance of the 10-bilayer Nb/TiN laminate calculated using an a genetic-algorithm based optimization tool; (2) the thermal resistance of the Nb/TiN laminate with thicknesses of 50 nm and 10 nm matching those in the literature (C. Bon-Mardion, et.al. 2024), but adjusted to 10-bilayer configuration to match the total thickness and the number of layers in our work; (3) a 10-bilayer Al/Re laminate (50 nm/10 nm), which was not optimized using the in-house optimization tool and (4) the thermal resistance of superconducting Nb bumps of the same length. As anticipated from the pronounced acoustic impedance contrast and the optimization of the coherent phonon scattering, the optimized Al/Re laminate consistently yields higher thermal resistance than both references across the entire sub-Kelvin regime.

This compact, sub-micron design is therefore compatible with the tight space constraints in flip-chip packaging. At the same time, its exceptionally high thermal resistance strongly suppresses heat flow into the device-carrying chip and preserves its lower temperature.

Moreover, to assess how deposition-induced thickness variations affect the thermal resistance of the Al/Re laminate, we performed a sensitivity study. For the given optimized Al/Re laminate, we generated a group of random realizations of laminates in which every layer thickness was perturbed by a small number $\varepsilon$. The thermal resistance of each variation was calculated and the mean and standard deviation of the ensemble were recorded. This procedure was repeated while increasing $\varepsilon$ from 1% to 50%. For randomized deviations from nominal layer thicknesses up to 10%, corresponding thermal resistance deviations remained below a few percentages. Increasing the upper limit of layer thickness deviations to 50% resulted in thermal resistance deviations below 30%. This result demonstrates that the Al/Re laminate's high thermal resistance performance is weakly sensitive to realistic deposition tolerances, indicative of its stability for scalable fabrication.

One of the main concerns with using superconductors as thermally insulative materials is the residual presence of Bogoliubov quasiparticles (electrons and electronlike holes) below $T_c$. While electrons freeze out exponentially below $T_c$, some marginal population of quasiparticles still exist in the superconducting state, and even small fractions of the normal-state population of electrons could carry heat more efficiently than phonons. These quasiparticles persist down to about 10-20% of $T_c$, which enforces a relatively low temperature limit for bilayers consisting of low $T_c$ materials such as Re ($T_c \approx 2.4K$) and Al ($T_c \approx 1.2K$). Furthermore, the superconducting gaps of each layer are also affected by neighboring superconductors (known as the proximity effect), which can significantly modify the quasiparticle densities. To study this effect, the Usadel equations were solved for the Re-Al optimized laminate to compute the superconducting gap profile across all the layers (See the Appendix D for details).

**Conclusions**

Our modeling and optimization framework found that layer thicknesses are crucial in constructing laminates that can block a broad band of phonons. A comparison between constant thicknesses versus varied thicknesses for 10 bilayers of Re/Al had a significant impact on the transmission rates, which corresponded to about two orders of magnitude increase in thermal resistance below 1K.

The optimization implementation also found that the material selection plays a key role in phonon filtering efficiency, especially when adding more bilayers to the stack. Some material combinations, such as Nb and TiN are great for filtering a large rate of phonons for a single interface, but do not exhibit any enhancement in their reflectivity when adding more interfaces. In contrast, materials such as Re and Al significantly benefit from increasing the number of bilayers, even though a single interface may not be an exceptional filter.

These results were both non-obvious and non-intuitive, given the complexity of the large wave effects in laminate systems consisting of up to twenty layers. In fact, traditional means of analyzing and comparing acoustic impedances between two materials was found to be a poor indicator of how well two materials can perform as phonon Bragg reflectors.

Ultimately, the methodology contained in this study could unlock new superlattice designs that are both thermally resistive and electrically superconductive. Future works could focus on optimizing with a material set that contains non-superconducting materials, larger layer thicknesses—as fabrication methods enable a wider range of thicknesses—and the proximity effect to better estimate the operating temperature range. This methodology could be applied for optimizing laminates for non-thermal phonons as well, to build qubit shields and other applications that require blocking a narrow band of phonons.

**Appendix A – Governing Equations**

The displacement field, $\boldsymbol{u_l} = u_l\hat{\boldsymbol{\imath}} + v_l\hat{\boldsymbol{\jmath}}$, and mechanical stresses, $\sigma_{xx,l}$ (normal) and $\sigma_{xy,l}$ (shear), associated with layer $l$ can be expressed by (A1) - (A4). Note that $e^{i(-k_y y-\omega t)}$ is dropped as a common factor across all terms ($k_y = k_0 \sin\theta_0$ is the same in all layers to enforce Snell's law; $k_0$ and $\theta_0$ are the wavenumber and angle of incidence of the wave injecting into the stack from the source material).

$$u_l(x) = cos\,\theta_{lL}\left(A_{lL}^{+}e^{i(k_{xL}x-k_{yL}y)} + A_{lL}^{-}e^{i(-k_{xL}x)}\right) - \sin\theta_T\left(A_{lT}^{+}e^{i(k_{xT}x-k_{yi}y)} + A_{lT}^{-}e^{i(-k_{xL}x)}\right) \quad \text{(A1)}$$

$$v_l(x) = \sin\theta_{lL}\left(A_{lL}^{+}e^{i(k_{xL}x)} + A_{lL}^{-}e^{i(-k_{xL}x)}\right) + cos\,\theta_T\left(A_{lT}^{+}e^{i(k_{xT}x-k_{yi}y)} - A_{lT}^{-}e^{i(-k_{xT}x)}\right) \quad \text{(A2)}$$

$$\begin{aligned}\sigma_{xx,l} = {} & \left(A_{lL}^{+}e^{i(k_{xL}x)} + A_{lL}^{-}e^{i(-k_{xL}x)}\right)\left[\lambda\left(ik_{xL}\cos\theta_{lL} + ik_y\sin\theta_{lL}\right) + i2\mu k_{xlL}\cos\theta_{lL}\right] \\ & + A_{lT}^{+}\left[\lambda\left(ik_{xT}\cos\theta_{lT} - ik_y\sin\theta_{lT}\right) + i2\mu k_{xlT}\cos\theta_{lT}\right]e^{i(k_{xT}x)} \\ & + A_{lT}^{-}\left[\lambda\left(-ik_{xT}\cos\theta_{lT} + ik_y\sin\theta_{lT}\right) - i2\mu k_{xlT}\cos\theta_{lT}\right]e^{i(-k_{xT}x)}\end{aligned} \quad \text{(A3)}$$

$$\begin{aligned}\sigma_{xy,l} = {} & A_{lL}^{+}\mu\left(-ik_{xL}\sin\theta_{lL} - ik_y\cos\theta_{lL}\right)e^{i(k_{xL}x)} + A_{lL}^{-}\mu\left(ik_{xL}\sin\theta_{lL}\, ik_y\cos\theta_{lL}\right)e^{i(-k_{xL}x)} \\ & + \left(A_{lT}^{+}e^{i(k_{xT}x)} + A_{lT}^{-}e^{i(-k_{xL}x)}\right)\left(ik_{xT}\sin\theta_{lT} - ik_y\cos\theta_{lT}\right)\end{aligned} \quad \text{(A4)}$$

The resulting system of equations could be solved either traditionally in a large matrix form, $\boldsymbol{h_{tot}A} = \boldsymbol{b}$, where $\boldsymbol{A} = [A_{0L}^{-}, A_{0T}^{-}, A_{1L}^{+}, A_{1T}^{+}, A_{1L}^{-}, A_{1T}^{-}, \ldots A_{NL}^{+}, A_{NT}^{+}]$ i.e., by $\boldsymbol{A} = \boldsymbol{h_{tot}^{-1}b}$ or with the transfer matrix method [15]. There are four unknown wave amplitudes per layer, and two unknowns in the source and sink materials for the total reflected and transmitted waves, respectively.

Calculating the quarter-wavelength thickness for a layer can be done with $\lambda_q = \frac{c}{4f}$.

**Appendix B – Genetic Algorithm**

The design space for laminates is quite large—for example, a laminate can be composed of twenty layers, each with a different layer thickness and material properties. This results in an optimization problem with a 20-parameter design space. Local optimization methods are not feasible for such a large design space, and therefore, to further improve the laminate design, a genetic algorithm was deployed in this study, utilizing the Non-dominating Sorted Genetic Algorithm II, known as NSGA-II, which was implemented in Python. A genetic algorithm could explore a large parameter space and its implementation in Python allows fine-tuning of various hyperparameters to improve the optimization search. For example, an adaptive mutation probability was added, which allows population-dependent genetic diversity to guide the mutation probability in each generation. The framework of the genetic algorithm was built based on the Differential Evolutionary Algorithms in Python (DEAP) library. At the same time, layer thicknesses of each layer were constrained to 50 nm based on fabrication guidelines.

The genetic algorithm is a nature-inspired optimization method that emulates biological evolution to find optima for complex functions. In this work, the Genetic Algorithm was implemented with the DEAP framework. To apply it to our laminate invention, we first defined the population and individuals. The population was sorted as a Python list. Each laminate was encoded by the thicknesses of its 20 layers, so an individual consisted of a 20-element vector of layer thickness values within prescribed bounds. Using DEAP's toolbox, we registered the functions that generate genes, built individuals and assembled the population. The fitness of individuals was obtained by evaluating the results from the CLAMM model. Since the problem was posed as a minimization, DEAP's fitness class was defined within a negative weight. Evolution proceeded generation-wise by mainly applying three operators (selection, crossover and adaptive mutation as described in the last paragraph). In each generation, the population was selected, recombined and mutated. The fitnesses of the offspring were evaluated using the CLAMM model, and the new generation replaced the previous one. This evolution was continued until the convergence criteria were satisfied. We chose a maximum of 200 generations based on trials showing rapid convergence within the first 50 generations. For simultaneously maximizing the thermal resistance and critical temperature while minimizing the total layer thicknesses of the laminate structure, the multi-objective optimization feature of NSGA-II was employed. In the evolution process, the population was first ranked by non-dominated levels, producing hierarchy fronts. When the size of the fronts exceeded the allowed population, the exceeded part was truncated using crowding distance. Binary tournament selection then chose individuals based on their ranks and thereby drove the search toward a Pareto front that offered the best trade-off among the three objectives.

Table 6. The material properties used in this study. Note that these material properties were approximated from known elastic properties or estimated based on existing literature.

| Material Name | $T_c\ (K)$ | $c_L\ \left(10^3\,\frac{m}{s}\right)$ | $c_T\ \left(10^3\,\frac{m}{s}\right)$ | $\rho\ \left(\frac{g}{cm^3}\right)$ |
|---|---|---|---|---|
| Al | 1.2 | 6.24 | 3.04 | 2.73 |
| Nb | 9.7 | 4.74 | 2.08 | 8.6 |
| NbN | 14 | 6.92 | 3.7657 | 8.2 |
| Re | 1.7 – 2.4 | 5.45 | 2.91 | 21 |

| Ta | 4.5 | 4.17 | 2.08 | 16.69 |
|---|---|---|---|---|
| TiN | 5.6 | 10.6 | 6.2 | 5.22 |
| V | 5.38 | 6.02 | 2.77 | 6.11 |
| Mo | 0.92 | 6.25 | 3.35 | 10.2 |
| Zr | 0.55 | 4.65 | 2.25 | 6.52 |
| Ti | 0.3 – 0.4 | 6.07 | 3.05 | 4.5 |

**Appendix C – Thermal Conductivity due to Quasiparticles**

An additional laminate design consideration is the superconducting transition temperature of the materials relative to the intended operating temperatures. Electronic thermal conduction may become dominant over phonon conduction in superconductors that are near, but still below, their superconducting transition temperature due to an ample population of heat-carrying electron/hole-like Bogoliubov quasiparticles. In such instances, the high lattice resistance offered by the laminate is effectively bypassed through electronic conduction across interfaces and electron-phonon coupling. A related recent work [Young et al. (manuscript in preparation)] demonstrated an accurate simulation of experimental measurements of the most extreme case of this effect: a normal metal to superconductor interface. This was accomplished by adapting an equation from Bardeen, Rickayzen, and Tewordt [25] to yield:

$$\left(R^{BRT}_{th,es}\, A\right)^{-1} = {G^{BRT}_{th,es}}/{A} \approx 4\left({k_B^2}/{e^2}\right)\left(\frac{T_c}{R_c^{BRT}A_c^{BRT}}\right) exp(-y)\left\{1+y+\tfrac{1}{2}y^2\right\} \quad \text{(A5)}$$

Here, $G^{BRT}_{th,es}/A$ is the interfacial thermal conductance [W/m$^2$-K] (the inverse of the interfacial thermal resistance $R^{BRT}_{th,es}\, A$ ), ${k_B}/{e}$ is Boltzmann's constant divided by the elementary charge, $y = \Delta/k_B T$, $\Delta$ the superconducting gap as a function of temperature, $T_c$ the superconducting transition temperature, and $R_c^{BRT}$ [Ω] is the electrical contact resistance across a contact area $A_c^{BRT}$ [m$^2$]. Accounting for this effect in a purely superconducting laminate is non-trivial as it requires consideration of the proximity effect between layers and quasiparticle tunneling between proximitized layers. In the case of a normal metal-superconductor interface, electronic thermal conduction generally becomes negligible around $\frac{T}{T_c} \approx 9$, but the purely superconducting laminate case will have reduced electronic conduction and therefore will maintain negligible electronic conduction at temperatures somewhat closer to the highest (non-proximitized) $T_c$ of the laminate.

**Appendix D – Usadel Equations**

Superlattice structures can be engineered to effectively block phonons—however, at mK temperatures electrons are the dominant carriers of heat. Superlattice materials have been chosen to be superconducting in the temperatures used for superconducting devices to minimize the amount of free quasiparticles (unpaired cooper pair electrons) at low temperatures. Superconducting logic families (RQL, AQFP, RSFQ, etc.) are all Nb based and are typically operated in liquid helium or using pulse tube refrigeration, at 4.2 K or 1.2-4.2 K respectively. Other use cases, involving sensors or qubits are typically operated at dilution refrigerator temperatures (low mKs) to minimized noise. We have examined superlattices in both temperature regimes to ensure that the structure does not have excess heat conduction of phonons. We calculate the gap and then examine the quasiparticle population available at different temperatures.

Modeling was conducted by solving the Usadel equations in COMSOL Multiphysics®. The Usadel equations are derived from two electron wavefunctions. From there one can use a quasiclassical approximation to go to a single particle function via integration. The dirty limit is the regime where elastic self-energy scattering dominates and allows one to apply the eilenberger equations. When the system is isotropic, we can use the diffusive Usadel equations.

The Ricatti equations were solved due to their numerical stability as a solution to these green functions. Since the Usadel equations are a mean field approximation the equations need to be solved self consistently. The two equations and the gap equations are shown below. The workflow consists of the equations being solved with the appropriate boundary conditions for a range of energies, with the gap solved for over all of these energies. The gap is then solved for these energies again until convergence is reached.

Like most FEA programs COMSOL® uses a Jacobian matrix (often called a stiffness matrix) to linearize and discretize nonlinear equations. The Jacobian must be inverted to solve and this can be costly depending on the condition number of the matrix.  The condition number of the unscaled equations was 10^6 and scaled they were. Edge cases were taken with the maximum and minimum energies evaluated at and the condition number was evaluated to determine the worst-case scenario. A scaling factor was applied to effectively lower the condition number. The following scaling factor was found to work across a range of energies, but in principle scaling factors could be applied at different energies to allow easier convergence. The results are shown in Figure D1 for the gap of the superlattice layers.

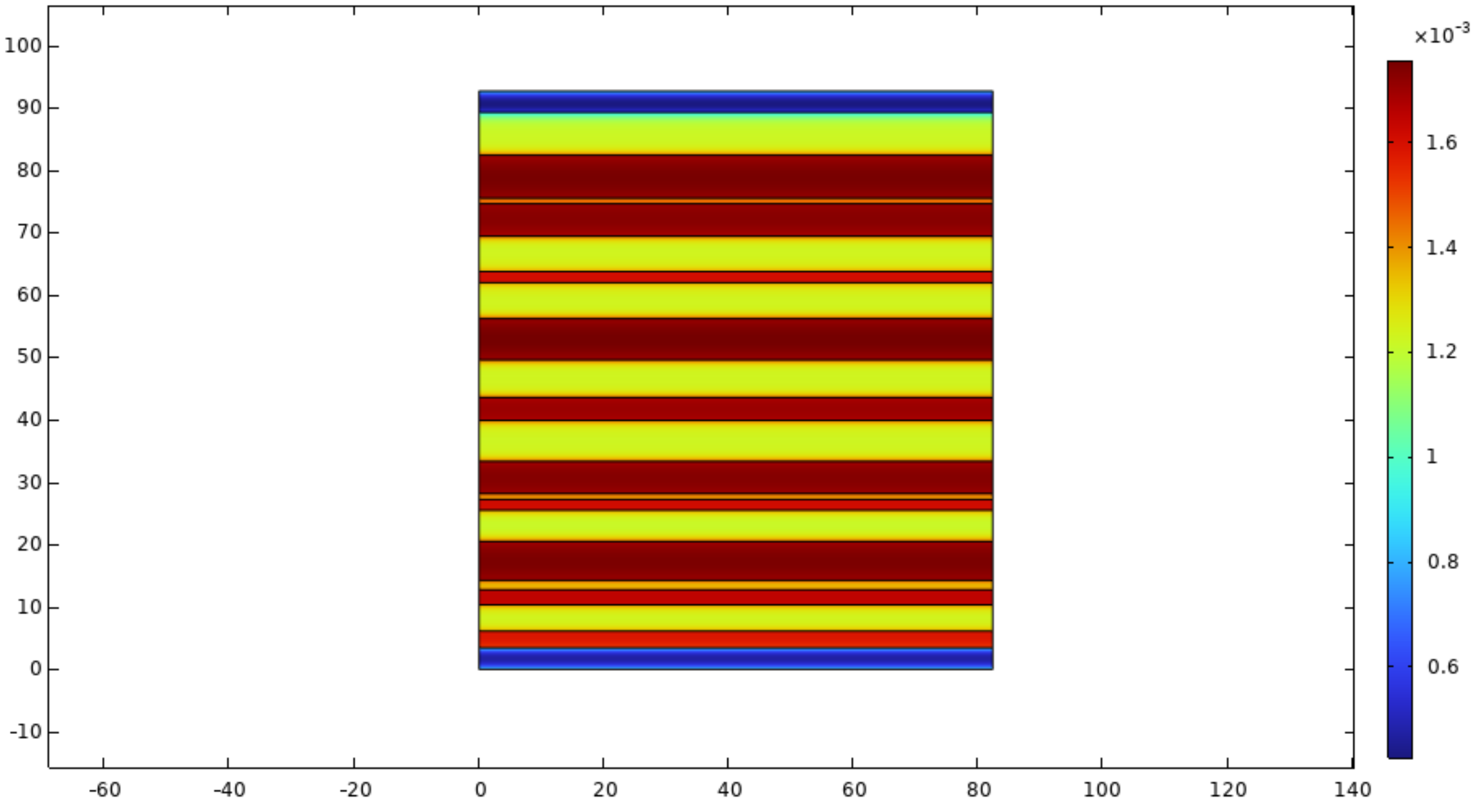

Figure D1. *Superconducting gap of the entire stack.*